\documentclass{pasa}
\usepackage{graphicx}

\title[Inner connection  for SNe Ic]{Toward an  inner connection  of SNe Ic, SLSNe Ic, XRF connected SNe,  SNe Ic-BL, and GRB connected SNe}

\author[Zou \& Cheng]{{Yuan-Chuan Zou$^{1,2}$, K. S. Cheng$^1$}\\
\affil{$^1$Department of Physics, Hong Kong University, Hong Kong, China}
\affil{$^2$School of Physics, Huazhong University of Science and Technology, Wuhan 430074, China; zouyc@hust.edu.cn(YCZ)}
}

\jid{PASA}
\doi{10.1017/pas.\the\year.xxx}
\jyear{\the\year}
\usepackage{aas_macros}
\usepackage{hyperref}
\usepackage[authoryear]{natbib}
\bibpunct{(}{)}{;}{a}{}{,}
\begin{document}

\begin{abstract}
Type Ic supernovae (SNe) can be classified as:  normal SNe Ic, type Ic super-luminous SNe (SLSNe Ic), X-ray flash (XRF) connected SNe,  broad-line SNe Ic (SN Ic-BL), and gamma-ray burst (GRB) connected SNe. Here we suggest an  inner connection   for all kinds of SNe Ic based on a pair of jets being successfully launched: a normal SN Ic is a normal core collapsar without jets launched; a GRB associated SN Ic is a core collapsar with relativistic jets launched and successfully breaking out the envelope of the progenitor; an XRF associated SN Ic is a core collapsar with jets launched but can only develop a relativistic shock breakout; an SN Ic-BL is an off-axis GRB or XRF associated supernova; and an SLSN Ic is close to the XRF-SN Ic but the shock breakout is not relativistic and most of the jet energy is deposited into the supernova component. Based on the luminosity-distance diagram, we derived the luminosity function of all different types of SNe Ic as a whole. We also show that the normal SNe Ic and GRB connected SNe Ic have similar accumulative distributions. 
\end{abstract}


\maketitle

\section{Introduction}

Gamma-ray bursts (GRBs) have been known for nearly fifty years \citep{1973ApJ...182L..85K}. Their properties and origins have been extensively studied \citep[see][for reviews]{2004RvMP...76.1143P,2006ARA&A..44..507W,2015PhR...561....1K}. Since the BATSE era (1990-2000), GRBs have been divided into two groups according to their duration and the spectral hardness ratio \citep{1993ApJ...413L.101K}. The short hard GRBs are believed to result from a black hole - neutron star (BH-NS) merger or NS-NS merger, while the long soft GRBs result from the collapsing of massive stars \citep{2004RvMP...76.1143P}.  It is clear that at least part of long GRBs are collapsars as they are connected with Type Ic Supernovae, as was discovered in the connection of GRB 980425 and SN 1998bw \citep{1998Natur.395..670G}, and then spectrally confirmed in the connection of GRB 030329 and SN 2003dh \citep{2003ApJ...591L..17S}. Hereafter, we use GRBs to denote long GRBs.

Type Ic supernova (SN Ic) is one of the final fates of a massive star \citep{1997ARA&A..35..309F, 2003ApJ...591..288H}.
Since the discovery of the connection between the long GRB and the SN Ic, the SN Ic can be divided into several sub-classes, which are: normal SN Ic, GRB associated SN Ic, X-ray flash (XRF) associated SN Ic \footnote{XRF is roughly taken as the burst in which the main prompt emission concentrates at hard X-rays rather than at $\gamma$-rays \citep{2003AIPC..662..244K}. There was also another subclass X-ray rich GRBs, which have more $\gamma$-rays than the XRFes have \citep{2005ApJ...620..355L}. However, there is no strict classification like the short and long GRBs. Here we also take the so called low luminosity GRBs as XRFes, as they are also soft, such as GRB 060218 \citep{2006Natur.442.1008C}, and GRB 100316D \citep{2011ApJ...740...41C}.}, broad line SN Ic (SN Ic-BL) and superluminous supernova Ic (SLSN Ic).

A unified picture is always a fundamental goal for the researchers, as it reveals a simple essence for the diverse phenomena, such as the unified models for different types of active galactic nuclei \citep{1993ARA&A..31..473A}.
\citet{2005IJMPA..20.6562D} even suggested a unified model for almost all the high energy phenomena by their ``cannon ball" model \citep{2004PhR...405..203D}.
\citet{2004ApJ...607L.103Y} provided a unified picture for different types of GRBs by counting the numbers of sub-shells of the jets: short GRBs with a few sub-shells,  long GRBs with a lot of sub-shells and X-ray flashes (and X-ray rich GRBs) being observed off-axis. However, one should notice that the essential difference between the long GRBs and short GRBs is the progenitor.
For the different types of GRBs based on their spectra, a unified picture has been suggested for XRFs, X-ray-rich GRBs, and GRBs by observing at different viewing angle of the same top-hat jet \citep{2005ApJ...620..355L}, or intrinsically non-uniform jet with either a power-law distribution \citep{2002MNRAS.332..945R} or a Gaussian distribution \citep{2004ApJ...601L.119Z}.

Different studies have been made to reveal the physics of the different types of SNe Ic, as well as their connection to the GRBs.
Just after the first detection of the association of GRB and SN (GRB 980425/SN 1998bw \citep{1998Natur.395..670G}),  \citet{1998ApJ...507L.131C}  suggested a unified scenario in which the explosion of SN also launches a pair of jets with opposite directions, which produces the GRB, and the asymmetry of the explosion provides a kick for the high speed pulsar.

Through the numerical simulation of a relativistic jet propagating through the stellar envelope,  \citet{2003ApJ...586..356Z} suggested that the X-ray flashes and the GRBs may be a consequence of different viewing angles, and the GRB associated supernova may be powered by the non-relativistic part of the jets. \citet{2012ApJ...750...68L} performed a hydrodynamic simulation by keeping the total energy budget as constant. They found that a longer lasting jet can overcome the envelope, while the shorter one may just power the SN component to be more energetic, and if it is short enough, the SN component cannot even be distinguished from a normal SN.
Most recently, \citet{2016arXiv160302350L} showed the GRB jets propagation through the progenitor envelope, depends largely on the parameters choosen.

\citet{2005ApJ...619..420L} suggested that the GRB jets are launched by the Blandford-Znajek process \citep{1977MNRAS.179..433B}, while the supernova component is launched by the spin energy transferred from the central black hole to the accreting disk in a magnetic coupling process \citep{1999Sci...284..115V,2011A&A...535L...6V}.

\citet{2006NCimB.121.1207N} tried to unify the GRBs and SNe by the mass of $^{56}$Ni, which depends on the rate of the energy deposited to form the $^{56}$Ni. They outlined the consequence of the deposition rate dropping as GRB connected with a hypernova, a less luminous supernova and without a supernova connection. When the deposition rate is smaller than a critical value $3\times 10^{51} {\rm erg \, s^{-1}}$, the process of nucleosynthesis changes dramatically and no considerable $^{56}$Ni is produced. After the discovery of  GRBs 060505 and 060614, which are GRBs with no SN component observations at a deep flux limit \citep{2006Natur.444.1050D,2006Natur.444.1047F}, in the same scenario,  with  numerical simulation, \citet{2007ApJ...657L..77T} showed that this kind of GRBs may have very little $^{56}$Ni been synthesized.

\citet{2008ApJ...687.1201K} found long GRBs and SNe Ic have similar locations in host galaxies. Both GRBs and type Ic SNe are located in similar star forming region.

{ After the discovery of SLSNe \citep[e.g.][]{2012Sci...337..927G}, people found the connection between them and GRBs. 111209A was observed connected with a possible SLSN SN 2011kl  \citep{2015Natur.523..189G,2016arXiv160606791K}. \citet{2013ApJ...778...67N} suggested blue supergiant model for the connection of super-long GRBs and SLSNe. \citet{2013ApJ...779..114V} proposed that SLSNe and long GRBs are both from young dense star clusters of their host galaxies. \citet{2017MNRAS.470..197Y} found the flares of them share similar empirical correlation between the luminosity and time-scale. \citet{2018MNRAS.475.2659M} suggested a misaligned magnetar model for magnetar thermalization and jet formation to connect   SLSNe and GRBs. \citet{2018arXiv180103312C} considered the spectral similarity among the host galaxies of the GRBs, SLSNe, star bursts and active galactic nuclei.}

\citet{2011ApJ...726...32F} suggested an alternated unified picture based on the central engine that: an energetic SN associated with a normal GRB   (e.g., GRB 030329/SN 2003dh \citep{2003ApJ...591L..17S}) comes from a powerful central engine plus immediate jets launching, an energetic SN associated with a low luminosity GRB (e.g., GRB 980425/SN 1998bw \citep{1998Natur.395..670G}) comes from a powerful central engine but with delayed jet launching from the envelope, and a less-energetic SN associated with a low luminosity GRB (e.g.,  XRF 060218/SN 2006aj \citep{2006Natur.442.1008C}) originates from an essentially less powerful central engine (slowly rotating magnetar).
On the other hand, \citet{2015ApJ...807..172N} tried to unify the low luminosity GRBs and normal long GRBs, by suggesting that they are both originated from a massive core collapsar, while the progenitor of the low luminosity GRBs has an extra extended envelope. This envelope prevents the jets penetrating and it results in a  low luminosity GRB.
They used this scenario to explain the association of the low luminosity GRB 060218/SN 2006aj.

Recently, \citet{2015ApJ...807..147W} suggested that the energy from the $^{56}$Ni and the rotational energy of the newly born neutron star may power the luminous SNe Ic, with different fractions for different SNe Ic. \citet{2016ApJ...826..178G} showed that the jet-feedback provides a variety of energy from the core collapsing star, which can power both the super-energetic supernova and the GRB jets. \citet{2017arXiv170408298P} showed the observational evidences for a possible connection between a normal core collapsing supernova and GRB by the choked jets.

\begin{figure}
 \centering
 \includegraphics[width=0.25\textwidth]{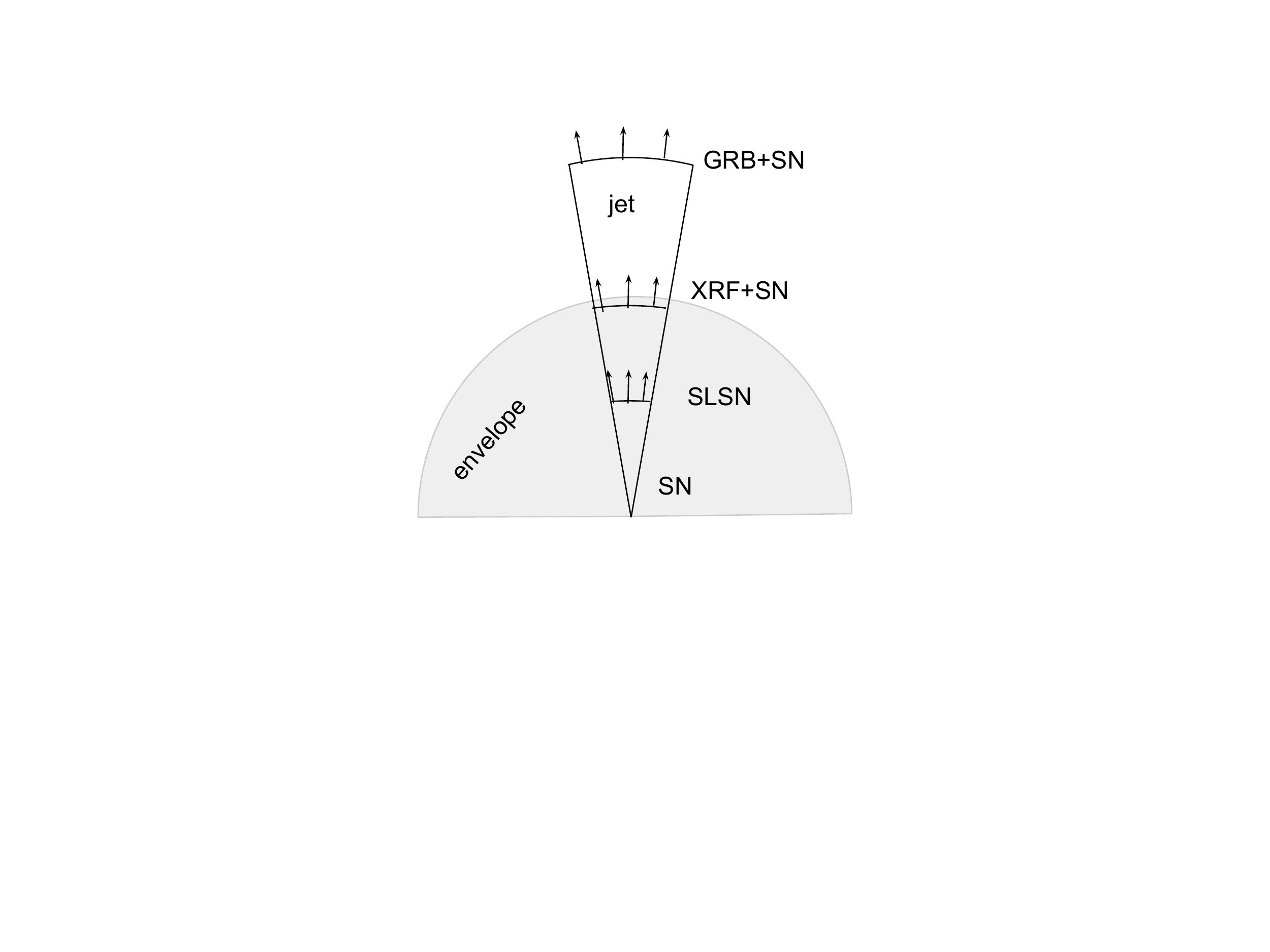}
 \caption{A diagram for the different core-collapsing SNe Ic without considering the effect of the line of sight. If the jet successfully passes through the envelope,  a GRB is associated with an SN Ic-BL. If the jet is blocked and the energy is released mainly as a relativistic shock breakout, an XRF is associated with the SN Ic-BL. If the jet is totally blocked, and most of the energy goes to the SN, it appears as an SLSN. While if there is not jet launched, it appears as a normal SN Ic.}
 \label{fig:sketch0}
\end{figure}

In the diagram of kinetic energy versus velocity of the ejecta, \citet{2014ApJ...797..107M} showed a continuous sequence between normal SNe Ic and GRBs/SNe, and the total  energy for both SNe  \citep{2014ApJ...797..107M}  and GRBs \citep{2003ApJ...594..674B}  covers a similar relatively narrow range ($\sim 10^{51}-10^{53}$ erg) \footnote{The `narrow' range comes from the fact that the isotropic equivalent $\gamma-$ray energy of GRBs is in the range of $10^{51}-10^{55}$ erg, e.g., Fig. 2. of  \citet{2013IJMPD..2230028A}. When the jet angle effect is considered, the real energy's range is much narrower. Notice that the total energy of GRBs consists mainly of $\gamma-$rays and kinetic energy, while  \citet{2003ApJ...594..674B}  just considered the $\gamma-$rays. People believe that the efficiency (energy of $\gamma-$rays divided by total energy) of GRBs is very high, up to 10\%-90\%. From the energy of $\gamma-$rays, we can infer that the total energy of GRBs could be of the same order as the energy of $\gamma-$rays or about one order of magnitude larger than it. However, the efficiency is still being debated. Some people do not believe the efficiency could be very high, because one cannot imagine a process to convert most of the kinetic energy into photons. On the other hand, as the observed energy of $\gamma-$rays is given, low efficiency means the total energy is very high, e.g., $10^{54}$ erg. Then one cannot imagine what kind of central engine can power such an energetic jet (See however, the Blandford$-$Znajek mechanism from a black hole central engine \citep{2013ApJ...765..125L}).  Therefore, unlike the supernovae, which can be inferred from the detailed observed light curves as well as the spectra, the precise amount of energy of GRBs is still in debate. The order of magnitude $10^{51}-10^{53}$ erg is reasonable. As our motivation is to figure out an  inner connection  of GRBs and different kinds of SNe Ic, we will not concentrate on the precise total energies.}. This indicates some inner connections in GRBs/SNe, as also suggested by \citet{2016ApJ...826..178G, 2017arXiv170408298P, 2017arXiv170401682Y}.
Motivated by the efforts on the unified scenario and the total energy of GRBs and SNe Ic are in the same order of magnitude, we suggest an  inner connection  for these different classes of SNe Ic, based on the behavior of the core-collapsing launched the pair of jets. A schematic diagram is shown in figure \ref{fig:sketch0}. 
If the jets with energy $\sim 10^{51}-10^{53}$ ergs are not launched by the central engine, it appears as a normal SN Ic with no broad emission lines.
The remaining classes occur when the launched jets ploughs into the envelope of the progenitor star.
If the jets successfully passes through the envelope, and one points  to the observer, it appears as a GRB, such as GRB 980425, while if it  does not point to the observer, we can only see an SN Ic-BL, and an off-axis afterglow is expected, such as the possible candidate SN 2002ap.
If the jets cannot successfully pass through the envelope, but most of the energy is transferred into the relativistic shock breakout, we can see it as an XRF, such as GRB 060218, while if most of the energy is deeply dissipated into the envelope, a major part of the energy is transferred into optical photons and appears as an SLSN.
\footnote{Very recently, when this paper is under reviewing, \citet{2017arXiv170500281S} suggested a similar picture about the  connection of GRBs and SNe Ib/c regarding whether the jets can be successfully launched, while they mainly concentrated on the normal SN Ib/c, low luminosity GRBs, and normal GRBs.}
We describe the scenario and its evidences  in section \ref{sec:body} and discuss it in \ref{sec:con}.

\section{Scenario and Evidences} \label{sec:body}
\begin{figure}
 \includegraphics[width=0.5\textwidth]{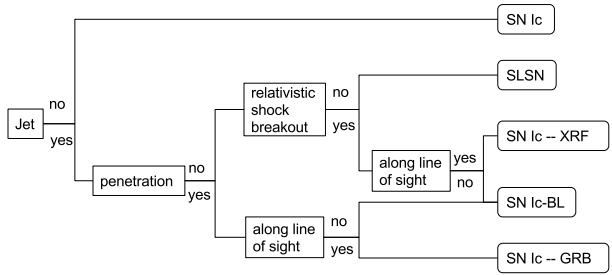}
 \caption{Schematic diagram for the scenario of different SNe Ic. Depending on whether there is a pair of jets, whether the jet successfully passes through the envelope, whether there is a strong shock breakout, and whether the jet is pointing to us, the observed type Ic supernova can be a normal SN Ic, an SLSN Ic, an SN Ic connected to an XRF, a broad line SN Ic, or an SN Ic connected to a GRB respectively}.
 \label{fig-sketch}
\end{figure}

As suggested by \citet{2012ApJ...749..110B}, the collapsar originated GRBs may also represent as short bursts, as the GRB jets may breakout of the envelope at different times. \footnote{The scenario  is supported from the view of GRB luminosity function very recently \citep{2017arXiv170701914P}.} It is also possible that the jets may not even overcome the envelope. However, the energy must come out, though it cannot come out as $\gamma$-rays. A very promising possibility is that the energy of the GRB jets is converted into optical photons, and results in an SLSN.
Numerical simulations also show that the relativistic jet may or may not successfully break the envelope \citep{2003ApJ...586..356Z,2006ARA&A..44..507W}.
Based on the fact that the core collapsar may or may not launch a pair of jets, and the jets may or may not penetrate the envelope, we propose the unified scenario as shown in figure \ref{fig-sketch}.

The full scenario is the following. The progenitor of SN Ic may or may not have a pair of jets. In the case of no jets, it becomes a normal SN Ic (e.g., SN 1994I). If it launches a pair of jets, there are two situations depending on whether the jets overcomes the envelope of the progenitor or not.  If the jets cannot pass through the envelope, the total energy of the jets is injected into the envelope. It becomes an SLSN (e.g., SN 2007bi). On the other hand, if some energy escapes as a shock breakout, one can see the XRF connected SN (e.g., SN 2006aj -- GRB 060218; SN 2008D -- XRF 080109). If the jet passes through the envelope successfully and is pointing to the Earth, we see a gamma-ray burst (e.g., SN 1998bw -- GRB 980425). On the other hand, if the jet is not pointing to the Earth, we only see an SN Ic-BL (e.g., SN 2002ap). Energy injection of the jets can create relative motion of particles in the envelope to produce broad line feature in the SN spectrum. This is a common result for SN Ic -- XRF, normal SN Ic-BL, and SN Ic -- GRB.

In the following subsections, we show evidences or indications for supporting this scenario.

\subsection{Luminosity distributions}
\begin{figure}
 \includegraphics[width=0.5\textwidth,angle=0]{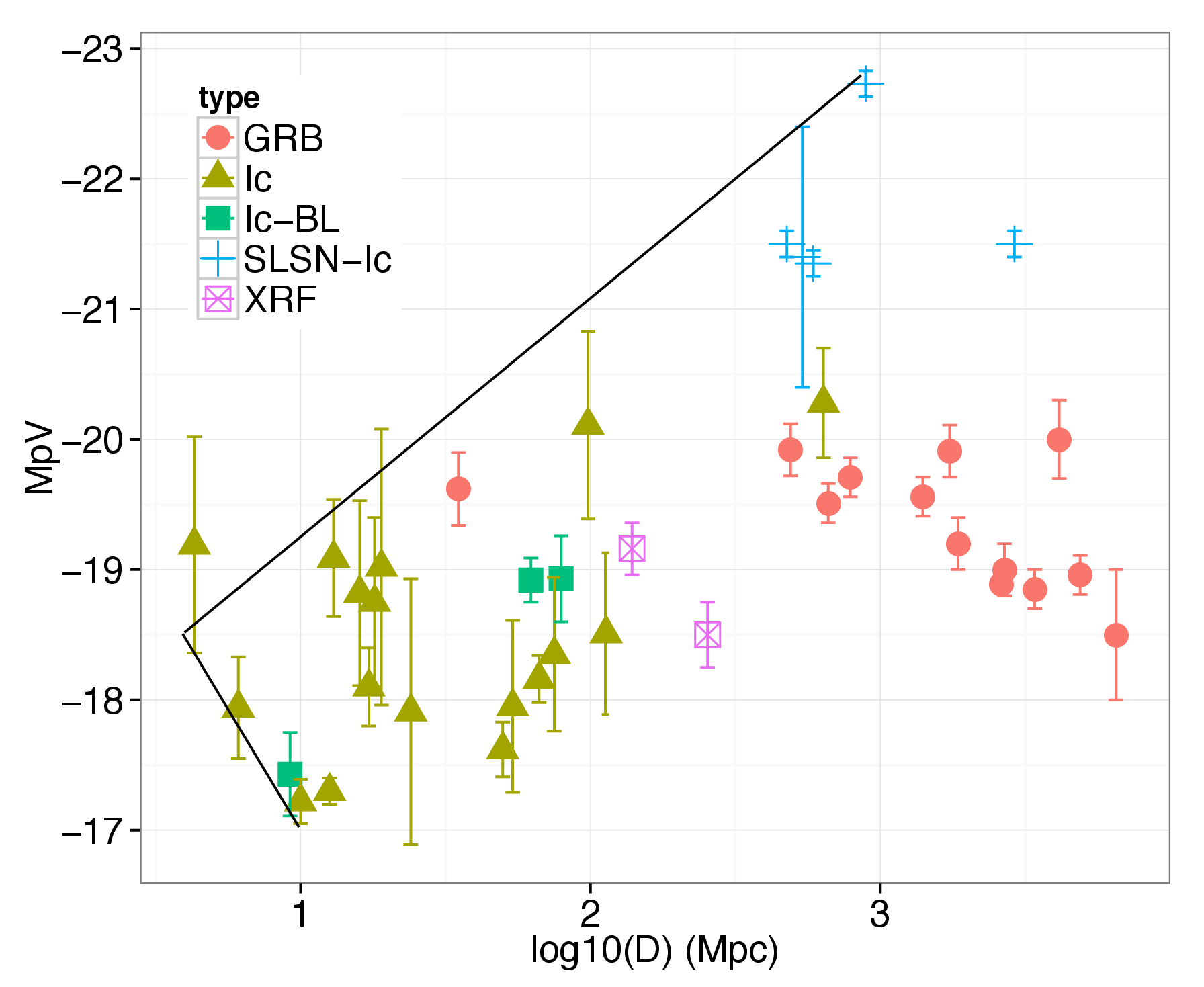}
 \caption{Luminosity-distance diagram \citep{2014AJ....147..118R} (V-band peak absolute magnitude of the SN component v.s. the luminosity distance in unit of Mpc) for all the SNe Ic, including normal SNe Ic (yellow triangles), SNe Ic-BL (green rectangles), SNe Ic associated with XRFs (marked as XRF, pink rectangles with crosses), SNe Ic associated with GRBs (marked as GRB, red dots),  and SLSN Ic (SLSN-Ic, blue crosses).  The two black solid lines show the edge of the data points, with slopes $\sim 2$ and $\sim -3$ respectively. The intersection locates at $\rm{M_{p,V}} \sim -18.6$.  Data are mainly from the  Asiago Supernova Catalogue (online updating data at http://cdsarc.u-strasbg.fr/viz-bin/Cat?B/sn) \citep{1999A&AS..139..531B},  and  a few other individuals are taken from literature which can be found in the references of the text.}
 \label{fig-Miller-dia}
\end{figure}

One piece of evidence for this scenario is the continuity among different types of SNe Ic in the luminosity space, which can be seen in the luminosity-distance diagram, as shown in figure \ref{fig-Miller-dia}.
The peak V-band absolute magnitude versus the luminosity distance is plotted for normal SNe Ic, SNe Ic-BL, SNe Ic/XRF, SNe Ic/GRB and SLSNe. On the bottom right of the panel, one can see a big empty region below the data of SNe Ic/GRB. This is a selection effect. For the normal SNe, they are not luminous enough to trigger the survey at far distances, while for the GRB connected SNe, the position is determined by the GRB satellites and follow-up afterglow monitoring, and then more powerful telescopes with longer observation time are used, so the much dimmer supernova component can be identified. Therefore, in this blank region, there should be more SNe Ic hidden. On the upper/lower left of the panel, the blank regions do mean there are no objects.

Because of the observational selection effect, one cannot derive the individual properties of the SNe population by directly counting the observed events. An alternative estimation could be performed based on the edge of the blank area on the upper left and bottom left of the panel, which shows roughly $L_{p,V} \propto D_{L,e}^{0.8}$, for $M_{p,V} < -18.6$ ($L_{p,V} > 8.7 \times 10^{42} {\rm erg\, s^{-1}}$), and  $L_{p,V} \propto D_{L,e}^{-1.2}$, for $M_{p,V} \geq -18.6$ ($L_{p,V} < 8.7 \times 10^{42} {\rm erg\, s^{-1}}$), where $L_{p,V}$ is the peak V-band luminosity directly from the absolute magnitude, and $D_{L,e}$ is the luminosity distance for the edge. Assuming a uniform space distribution of the SNe, the $D_{L,e}$ means the average distance between two SNe with the same luminosity, i.e., $D_{L,e} \sim [\frac{1}{n(L)}]^{1/3}$, where $n(L)$ is the number density at a certain luminosity. Therefore, we can get a rough luminosity function for all the SNe Ic \citep{2017arXiv170103261Z}:
\begin{eqnarray}
  n(L) \propto \left\{
  \begin{array}{lr}
  L_{p,V}^{2.5}, & L_{p,V} < 8.7  \times 10^{42} {\rm erg\, s^{-1}}, \\
  L_{p,V}^{-3.75}, & L_{p,V} \geq  8.7 \times 10^{42} {\rm erg\, s^{-1}},
  \end{array}
  \right.
  \label{eq:n_L}
\end{eqnarray}
which is roughly similar to the luminosity function of SNe Ia \citep{2010AJ....139...39Y} while different from that of GRBs \citep{2015ApJS..218...13Y}.
In comparing with luminosity-distance diagram for all kinds of SNe shown in figure 2 of \citet{2006AJ....131.2233R}, one can see no  edge on either the upper or lower left area. It is simply because different types of SNe obey  different luminosity functions. The edge as clearly shown in our figure \ref{fig-Miller-dia} suggests that different types of SNe Ic are under the same population in some sense.

Concentrating on the peak magnitude  in figure \ref{fig-Miller-dia}, one can see, roughly, that the luminosity is increasing in the sequence: normal SNe Ic, SNe Ic-BL \footnote{As all the SNe Ic associated with GRB or XRF, are also spectrally similar to a normal SNe Ic-BL \citep{2015PhR...561....1K}, here we take SN Ic-BL, SN Ic/XRF and SN Ic/GRB as one type.}, and SLSN. They are overlapping, while a K-S test also shows that the SNe Ic-BL is intrinsically brighter than the normal SNe Ic ($M_{\rm p,V,Ic}=-18.0\pm 0.5$, $M_{\rm p,V,Ic-BL}=-18.3\pm 0.8$ for V band, and $M_{\rm p,R,Ic}=-18.3\pm 0.6$, $M_{\rm p,V,Ic-BL}=-19.0\pm 1.1$ for R band) \citep{2011ApJ...741...97D}. For the normal SNe Ic, there is no jet and consequently no extra energy. Therefore, the luminosity is the lowest. For the SNe Ic-BL, they are either associated with on-axis GRB/XRF, or off-axis GRB/XRF (no GRB/XRF observed). Some extra energy from the jets goes into the progenitor envelope. Therefore, the luminosity of the SN component is brighter. For the SLSNe, most energy of the jets is deposited into the envelope. Therefore, the luminosity is the highest. Notice the intrinsic variety of the SNe and the jets, the difference of the SNe Ic is not strictly departed.

From the point of view of energy, SN-XRF, SN-GRB and   SLSN all have similar amounts of total energy. Normal SN Ic and SN Ic-BL should have smaller total energy because of no jet component or jet component is pointing to other direction. As the peak luminosity of the SLSN reaches more than $10^{44}$ erg/s, and peaks at around 100 days, the total radiation energy reaches to $\sim 10^{51}$ erg \citep{2012Sci...337..927G}. The energy of the GRBs is also of the order of $10^{51}$ erg \footnote{The observed $\gamma$-ray energy is around $5\times 10^{50}$ erg \citep{2001ApJ...562L..55F}. Considering the efficiency converting total jet energy into $\gamma$-ray energy being around 0.2 \citep{2015PhR...561....1K}, the total energy of the GRB jets is around $2.5 \times 10^{51}$ erg.}. 
The inferred kinetic energy of a normal SN Ic is $\sim 10^{51}$ erg, and it is $\sim 10^{52}$ erg for an SN Ic-BL \citep{2011ApJ...741...97D}.
 These figures are consistent with the  inner connection, that roughly speaking, if the jet energy is less than  $\sim 10^{51}$ erg, there is no observational effect and it appears as a normal SN Ic, while if the jet energy is larger than this value, it appears as an SLSN or a GRB etc.
Notice again, they should also have some diversity because of the diversity of the intrinsic energy reservoirs, which could be any of the GRB central engines, such as a black hole accretion disk system or a magnetar.




\subsection{SN Ic -- GRB}
As the GRB comes from the relativistic jet, for the GRB connected SN Ic, it has to produce a pair of jets. One directed to the observer, and the other in the opposite direction. The spectra of many events have been confirmed, such as GRB 980425/SN 1998bw \citep{1998Natur.395..670G}, GRB 030329/SN 2003dh \citep{2003ApJ...591L..17S}, GRB 031203/SN 2003lw \citep{2004ApJ...609L...5M,2004A&A...419L..21T},
GRB 091127/SN 2009nz \citep{2010ApJ...718L.150C},
GRB 101219B/SN 2010ma \citep{2011ApJ...735L..24S},
GRB 120422A/SN 2012bz \citep{2012A&A...547A..82M},
GRB 130427A/SN 2013cq \citep{2013ApJ...776...98X},
and GRB 130702A/SN 2013dx \citep{2015A&A...577A.116D}.
 \citet{2015Natur.523..189G} suggested that GRB 111209A/SN 2011kl  is an SLSN though its peak magnitude is -20 and it is dimmer than the usual SLSNe. It may come form a sub-group as a  ``blue supergiant" \citep{2013ApJ...778...67N}.
There might be a link between SLSNe and GRB connected SNe.
By accumulating more and more samples, one can tell what fraction of the long GRBs are associated with SNe Ic, and what fraction of the  SNe Ic are associated with GRBs \citep{2007ApJ...657L..73G}. 
\citet{2004ApJ...607L..13S} suggested that less than 6\% of the SNe Ic produces relativistic jets by searching  for radio emission from misaligned jets.
This may reveal the nature that which properties are crucial for producing a pair of relativistic jets.

\begin{figure}
 \includegraphics[width=0.5\textwidth]{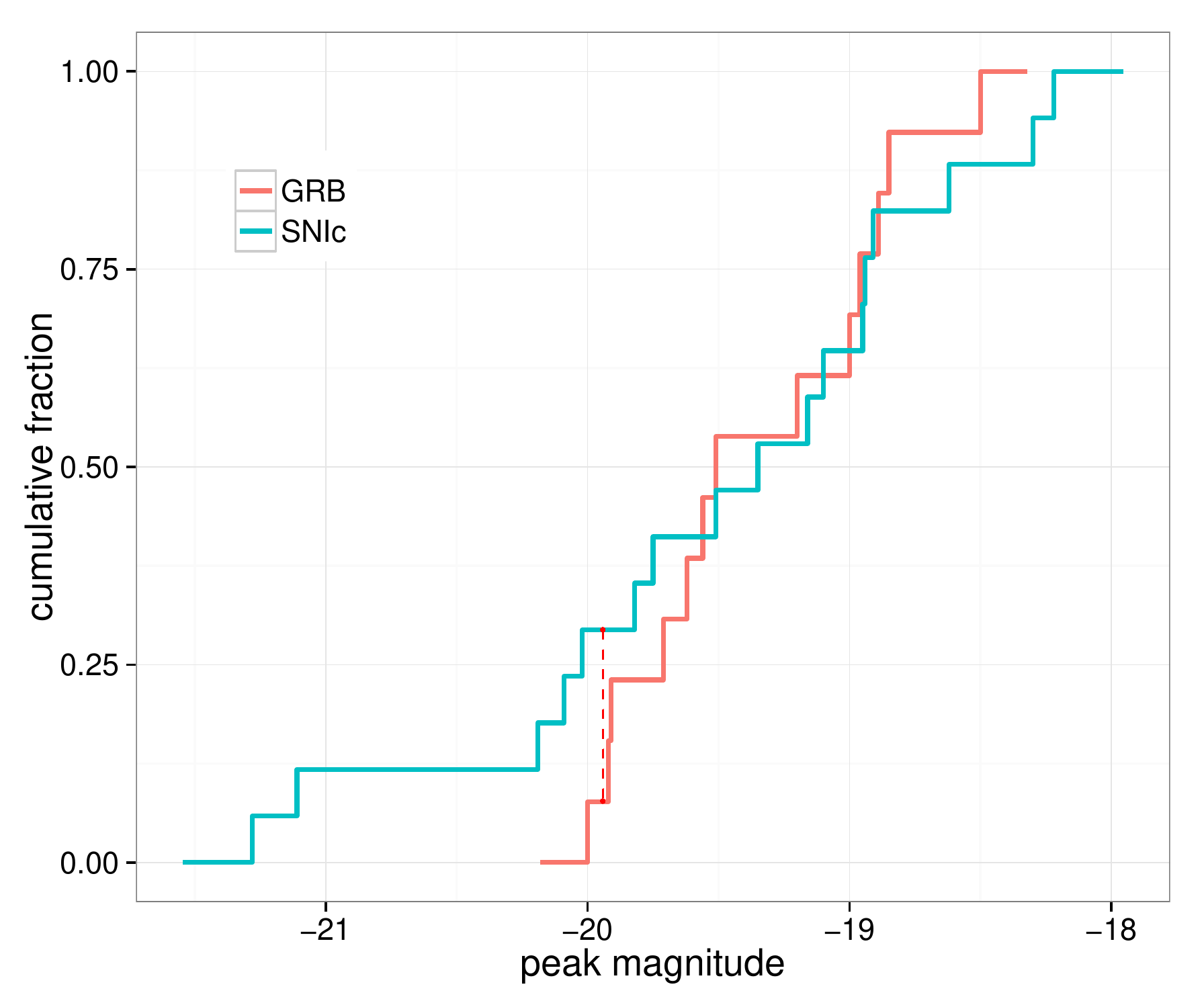}
 \caption{Kolmogorov-Smirnov test for the peak magnitude of GRB associated SNe (red) and normal SNe Ic (cyan), of which the GRB associated SNe are increased by  1 magnitude. The red dashed line indicates the maximum difference. Data are taken from the same source as in figure \ref{fig-Miller-dia}. }
 \label{fig-ks-GRB-SN}
\end{figure}

Figure  \ref{fig-ks-GRB-SN}  shows the Kolmogorov-Smirnov test for the peak magnitude of GRB associated SNe and normal SNe Ic. 
{ Bloom (2003) \citep[Figure 10 in ][]{2004RvMP...76.1143P} plotted the comparison between the GRB associated SNe and the local normal SNe Ib/c. The distributions were quite different. However, the data were very limited. Similar cumulative plots have been shown in \citet{2012grb..book..169H,2013RSPTA.37120275H}, and the relatively brighter peak luminosities for GRB associated SNe were suggested to be the result of a bias against the faint systems. \citet{2011MNRAS.413..669C} found the distributions are similar by neglecting those samples not having host galaxy extinction information.}
{ Here we assume the sample is complete, as some of the GRB afterglows have been observed in a deep flux, such as GRBs 060605 and 060614 \citep{2006Natur.444.1047F}. We increased the peak magnitude of GRB associated SNe by 1 Mag,  to look for any similarity between these two samples.} It shows that these two samples may share a similar distribution, which implies a similar origin, e.g., a similar range of the massive star.
A possible explanation might be that the supernova component is dominated by the mass of the progenitor, while the jet is dominated by the rotation of the progenitor, and the rotation is not related to the mass. Therefore, the supernova with or without a jet shows a similar distribution.
Notice the magnitude of GRB associated SNe are increased by 1 magnitude, which means they are relatively brighter than the normal SNe Ic. The reason might be that, for the supernovae with jets, part of the jet energy may deposit into the envelope, finally ending up as the optical luminosity.
However this constant off-set is likely observational bias. The normal SN Ic in the distance of GRB associated SNe will not be found by the routine survey. Once they are associated with GRB, an extensive search will be carried so apparent dimmer but intrinsic more powerful SNe can be detected. Furthermore, the beaming effect of GRBs make them difficult to detect in the smaller nearby volume but more easily detected at larger cosmological distances. As a conclusion, the similar distribution of the magnitude of the normal SNe Ic  and the modified magnitude (1 magnitude added) of the SNe Ic with GRB connection is interesting, but the reason might be complex and the similarity might be a coincidence.

\subsection{SN Ic -- XRF}
It is possible that the jet cannot fully penetrate the envelope, either because the envelope is too thick or the energy of the jet is relatively low. However, the jet is still energetic enough to power a relativistic shock breakout \citep[see][for an extensive modeling]{2010ApJ...725..904N,2012ApJ...747...88N}.
This breakout may radiate thermal X-rays, which can be detected as an XRF or an X-ray rich GRB (or presents as a low luminosity GRB \citep{2015MNRAS.448..417B}). The light curve of the XRF is smooth compared with a normal GRB, and the spectrum is thermal rather than non-thermal.
Several XRF/SN systems have been observed, such as
GRB 060218/SN 2006aj \citep{2006Natur.442.1008C,2006Natur.442.1011P,2006Natur.442.1014S},
and GRB 100316D/SN 2010bh \citep{2011ApJ...740...41C,2012ApJ...753...67B}.
This phenomenon has attracted greatest interests, and is successfully explained by a shock breakout model with some tuning or modifications \citep{2006Natur.442.1008C,2007ApJ...667..351W,2007MNRAS.375..240L,2015ApJ...807..172N} (see \citet{2016MNRAS.460.1680I} however for an alternative jet model).

The off-axis case has probably been observed already, such as SN 2007gr \citep{2010Natur.463..516P}, and SN  2009bb \citep{2010Natur.463..513S}. For SN 2007gr, a mildly relativistic jet directed far away from the observer was suggested \citep{2010Natur.463..516P, 2011ApJ...735....3X} \footnote{See however, \citet{2010ApJ...725..922S}  suggested SN 2007gr is just a normal SN Ic without a pair of mildly relativistic jets, and \citet{2011ApJ...728...14P} also argue that SN 209bb should not have a pair of relativistic jets.}.

SN 2008D \citep{2008Sci...321.1185M} is an SN Ib with a shock breakout \citep{2008ApJ...683L.135C}, which is connected to XRF 080109 \footnote{It is also noted as an X-ray transient (XRT) \citep[e.g., in][]{2008ApJ...683L.135C}.}. This may be a link among SNe Ib, SNe Ic and GRBs. For SNe Ib, as the existence of helium envelopes, the jets, if they exist, penetrate more hardly. It can only appear as an XRF rather than a normal GRB.


\subsection{SN Ic - BL}
Besides powering the relativistic jets, the central engine may also provide energy to the envelope of the progenitor. This extra energy may result in the ejecta of the supernova having a higher speed compared to normal supernova without a pair of jets. This is why all the SNe Ic connected with GRBs/XRFes are broad line SNe. 
GRBs are highly beaming events. That means a great part of GRBs are not observed because of the direction of the jet, while the percentage depends on the jet angle.
However, the beaming effect for the supernova component is negligible. Therefore, an off-axis GRB jets cannot be observed as a GRB, but the supernova can still present as a broad line SN. A similar effect applies to XRF connected SN Ic. However, as there is no relativistic jet propagating into the surrounding medium, one cannot expect an orphan afterglow.

Several SNe-BL have been observed, such as
SN 2002ap \citep{2002ApJ...577L..97K},
SN 2003jd \citep{2005Sci...308.1284M},
SN 2010ah \citep{2013MNRAS.432.2463M},
SN 2012ap \citep{2014ApJ...797..107M},
and PTF10qts \citep{2014MNRAS.442.2768W}.
People have paid high attention to the probability that these events are also off beamed GRBs, such as \citet[][for SN 2002ap]{2002ApJ...580L..39K,2003ApJ...598.1151T}, and other SNe-BL \citep{2006ApJ...638..930S}. However, there is no strong evidence for the existence of relativistic jets, though \citet{2005Sci...308.1284M} suggested SN 2003jd is an SN with off-axis GRB jets from the double-peaked emission line.
Recently, \citet{2015arXiv151201303C} searched the radio emission from the SN Ic-BL trying to find a signal from the off-axis jet, setting a probability  $\le$45\% of being associated with a GRB.
This scenario can be confirmed if an off-axis emission of a relativistic jet associated with an SN Ic-BL is detected, either in the optical band \citep{2002ApJ...579..699N,2007A&A...461..115Z}, or in  the radio band \citep{2002ApJ...576..923L}.
The  reason for the missing optical and radio signals might be due to the high beaming effect of the jet from a GRB, or even the absence of a jet from an XRF.
For example, based on the similarity between SN 2003jd and SN 2006aj,  \citet{2008MNRAS.383.1485V} suggested SN 2003jd may also be connected with an XRF.

The birth rate of the SN Ic-BL should be related to the SNe Ic connected with GRBs/XRFes, as these are just geometrical effects.
Lacking data, here we apply a very rough  estimation similar to that used in eq. (\ref{eq:n_L}), i.e., we take the distance $D$ of the nearest object as the average distance between two objects, and get the rate $\propto D^{-3}$.
As shown in Fig. {\ref{fig-Miller-dia}}, the nearest SN Ic-BL is 9.2 Mpc for SN 2002ap \citep{1999A&AS..139..531B}, and the nearest SN Ic -- GRB is 35 Mpc for SN 1998bw \citep{1998Natur.395..670G}. (The nearest SN Ic -- XRF is 139 Mpc for SN 206aj \citep{2006Natur.442.1008C}, much farther than the nearest SN Ic -- GRB. Hence, the birth rate of  SN Ic -- XRF is negligible comparing with the birth rate of SN Ic -- GRB.) A comparison of the birth rate of the SNe Ic -- GRBs and the SNe Ic-BL is $\sim (9.2 {\rm Mpc})^3/(35 {\rm Mpc})^3 = 0.018$. Considering the GRBs are beamed because of the relativistic jet, while SNe Ic-BL can be seen in any direction, this rate indicates that the solid angle of the jet (with half opening angle $\theta_j$): $\frac{\theta_j^2}{2} \sim 0.018$. 
This number is consistent with the estimation of the solid angle being $10^{-3}-10^{-2}$ from the jet opening angle \citep{2015PhR...561....1K}.

\subsection{SLSN Ic}
If the envelope of the progenitor star is too massive, the jets can never overcome the envelope. Then all the energy of the jet will be deposited into the envelope.
The mass-radius ($M-R$) relation of a WR star is given by \citet{1992A&A...263..129S},
\begin{equation}
\log \frac{R}{R_\odot} = -0.6629 + 0.5840 \log \frac{M}{M_\odot},
\end{equation}
where ${R_\odot}$ and ${M_\odot}$ are the radius and mass of the Sun respectively. Taking SLSN 2007bi as an example, the inferred mass the WR star  of SN 2007bi is $\simeq 43 M_\odot$ \citep{2010ApJ...717L..83M}. Consequently, the radius is $\sim 2 R_\odot$. Notice the shocked envelope propagates in mildly relativistic velocity \citep{2003ApJ...586..356Z,2004ApJ...608..365Z}. By taking $0.1 c$ as the velocity, it takes the jet $\sim 50$ s to pass through the envelope. Roughly we can say that only if the central engine last longer than $\sim 50$ s, it can support the jet to pass through the envelope. The typical time scale of the long GRBs is $\sim 26$ s \citep{1993ApJ...413L.101K}, which can marginally fail to support the jet in overcoming the envelope. Therefore, most of the jet energy is deposited into the envelope. On the other hand, the WR star mass of the SLSN is about ten times larger than that of the GRB progenitor star, which also makes it reasonable that the jets cannot pass through the envelope.
In Figure \ref{fig-Miller-dia} it appears that SLSN Ic have a larger intrinsic power than those GRB connected SN Ic. However as we have emphasized in our picture that GRB connected SN Ic are those with  a jet penetrating the envelope whereas SLSN Ic are those where the jet energy is deposited in the envelope. If we take into account the GRB energy in the GRB connected SN Ic, the combined energy is actually similar to that of SLSN Ic. 

{ The mass of the initial jet from the central engine is the same as the jet from a normal GRB, of which the total energy is $10^{51}-10^{52}$ erg, the initial Lorentz factor of the jet is $\sim 100$, and consequently the mass is about $0.01 {\rm M_{\odot}}$ \citep[see][for example]{2001ApJ...550..410M}. With the propagation of the jet inside the envelope, more materials are accumulated onto the jet and it can be slowed down to about $0.1$c. Considering the total energy is fully deposited into the kinetic energy of the envelope, the accumulated mass of the jet is consequently about $0.05-0.5 M_{\odot}$. Considering the extreme case of SN 2007bi, the kinetic energy was suggested being $3.6 \times 10^{52}$ erg \citep{2010ApJ...717L..83M}, the corresponding mass of the pair of the jet is roughly $ 2 M_{\odot}$, if the velocity is taken as $0.1 {\rm c}$. Taking the full mass of the envelope as $40 M_{\odot}$ as suggested in  \citet{2010ApJ...717L..83M}, which indicates the opening solid angle of the jet  is roughly $\frac{1}{40}$. However, if the kinetic energy is deposited into a wider angle in the envelope, the accumulated mass is heavier and the final average velocity is smaller.}

The energy from the jets may enhance the luminosity of the supernova dramatically, either by nucleosynthesis in terms of storing the energy in $^{56}$Ni \citep[e.g.][]{1982ApJ...253..785A,2013ApJ...773...76C}, or by being stored as thermal energy of the envelope \citep[e.g.][]{2009ApJ...696..953C}, or by transferring into kinetic energy of the envelope and then interacting with the circumstellar medium (CSM) \citep[e.g.][]{2011MNRAS.415..199M,2013ApJ...773...76C}. Together with the fact that the massive star itself may produce a brighter supernova, the final result may present as an SLSN. However, the observations make it difficult to distinguish between the Ib and Ic. The SLSNe are divided into three classes, i.e., radioactively powered SLSN-R, hydrogen-poor SLSN-I and hydrogen-rich SLSN-II \citep{2012Sci...337..927G}, where SLSN-R is also type Ic.
Recently, there have been several SLSN-R detected,
such as
SN 1999as \citep{2012Sci...337..927G},
SN 2007bi \citep{2009Natur.462..624G},
PTF12dam \citep{2013Natur.502..346N},
PS1-11ap \citep{2014MNRAS.437..656M},
and SN 2015bn \citep{2016arXiv160304748N}.
In figure \ref{fig-Miller-dia}, we only take SLSN-R into account.
There are three leading models:  magnetar spin down, $^{56}$Ni decay, and ejecta-CSM interaction \citep{2013ApJ...773...76C}. For the last two models, the central engine could be the energy injection from the GRB jets.

Recently, a very luminous SLSN ASASSN-15lh has been discovered with a peak bolometric luminosity $(2.2 \pm 0.2) \times 10^{45} {\rm erg\, s^{-1}}$ \citep{2016Sci...351..257D,2015ApJ...807..147W}. The magnetar model can hardly provide so much energy, and a quark star model has to be invoked \citep{2016ApJ...817..132D}. On the other hand, there is no such energy limit for a jet driving a SLSN. Very recently, \citet{2016arXiv160504925C} found the metallicity of the progenitors of the SLSN is relatively lower. This is consistent with the low metallicity stars having more extensive envelopes.

The total radiated energy of SLSNe is of the order of $10^{51}$ erg \citep{2012Sci...337..927G}. If they are powered by the relativistic jets, the energy should be comparable with GRBs, for which the radiated energy is truly also of the same order of $10^{51}$ erg \citep{2001ApJ...562L..55F}.  Here we assume the unknown radiation efficiency from jet energy to $\gamma$-rays of GRBs is of the same order as that of the jet energy to optical photons of SNe.  Considering the range of the jet energy, which varies from $1.1\times 10^{50} $ to $ 3.3 \times 10^{53}$ erg \citep{2011ApJ...734...58S}, we predict that, with accumulating data of SLSNe, there should be a fraction with energy up to $10^{53}$ erg.

\subsection{Normal SN Ic}
A type Ic SN is both H and He poor in the spectrum, which indicates the progenitor is a very massive star \citep{2009ARA&A..47...63S}.
The peak absolute magnitude is about -18, and the kinetic energy is about $10^{51}$ erg \citep{2011ApJ...741...97D}.
There is no evidence that a normal SN Ic has any jet. \citet{2003ApJ...599..408B} carried out a radio search of 33 SNe Ib and Ic observed from  1999 to 2002, and none of them showed clear jet emission. It is also possible that there is a jet as in the SLSN case, and the total energy of the jet is much less than $10^{51}$ ergs. Consequently, the deposited energy has no observational effect comparable to a normal SN Ic.


\citet{2009ARA&A..47...63S} showed a picture where a  fast rotating WR star may produce an SN Ibc-BL, while a slow rotating WR star will produce a normal SN. This is consistent with our picture, as a rotating WR star may also launch a pair of jets.


\section{Conclusions and Discussions} \label{sec:con}

Relating two or three sub-classes of SNe Ic has been suggested  widely in the literatures. In this paper we suggested an inner connection for all different sub-classes of SNe Ic. It is mainly based on the fate of the central engine launched jets. Roughly speaking, the more energy of the jets that is deposited into the envelope of the progenitor, the stronger the luminosity of the SN component. In addition to collecting previous precise evidence support our full picture, we also find new evidences in support of this inner connection.  Figure \ref{fig-sketch} schematically illustrates a full picture of the inner connection. We also propose a luminosity-distance diagram and use it to indicate the existence of a single luminosity function for all different types of SNe Ic, which obeys some inner connection. We estimate that the intrinsic power of SLSN Ic and GRB associate SN Ic have similar intrinsic power if the energy of the GRB is included. This evidence supports our picture that the difference between SLSN and GRB associated SNe is due to the absorption of the jet energy. As there is a small part of the GRBs which are much more energetic, we also predicted that much more luminous SLSNe will be observed with accumulating data.

This full picture also shows a hint why the GRBs only associate with type Ic SNe. The type II core collapsing SNe may also launch a pair of jets. The critical parameters to launch jets might be the mass of the initial main sequence star, its magnetic field, spin etc. But the most important factor is that for the type II SNe, they should have a more extensive envelope due to the hydrogen shell. This prevents the jets penetrating the envelope, { and consequently it cannot produce a GRB}. Maybe XRF 080109 is an example that the jets are launched from a type II SN, but have not penetrated. The other hint is the SLSNe with type II SNe. If the jets cannot pass through the envelope, but the energy is high enough, they may still power an SLSN, which is type II.




More detailed models should consider the extent of the energy budget for individual events. Failed jets may produce SLSN. While if it is not so energetic, it may produce a normal  SN Ic-BL. Therefore, the type SN Ic-BL may not only be associated with off beamed GRBs/XRFes.  
If the energy of the jets is even weaker, say  $\ll 10^{51}$ erg, the jets may not even penetrate a normal SN Ic progenitor. Because of the jet energy is negligible, there is no obvious observational evidence, and they are indistinguishable from an SN Ic with no jets launched.
The inner engine could be sufficiently energetic to penetrate an extensive envelope.
It is possible that part of the energy of the jets penetrates into the envelope, while the jets still pass through it. Consequently, the connected supernova is  brighter than a normal supernova, but dimmer than an SLSN, as seen in SN 2011kl (connection with GRB 111209A) \citep{2015Natur.523..189G,2016arXiv160606791K}.
The progenitor of GRB 111209A is also suggested as a blue supergiant by \citet{2013ApJ...766...30G}.
These might be minorities.




We also notice there are some long GRBs, which have been observed in a deep flux limit but the supernova component did not appear, such as GRB 060605, GRB 060614 \citep{2006Natur.444.1047F}, 	and
GRB 100418A \citep{2012PASJ...64..115N}. This is either because the associated supernova is essentially low luminosity \citep{2009Natur.459..674V}, or the GRB belongs with other sub-classes, such as, a merger originated GRB \citep{2007ApJ...655L..25Z},  an intermediate-mass black hole powered GRB \citep{2010ApJ...717..268G}, or a macronova associated GRB \citep{2015NatCo...6E7323Y}. According to these explanations, they are very likely not correlated with a core-collapsing supernova. { Future gravitational wave detection should be a key discriminator for the single or double origins of this kind of long GRBs.}
An alternative possibility is that, the progenitor star collapses directly into a black hole rather than a neutron star first. With no neutron star stage, the supernova cannot form, while only jet launched GRB appears.

It is unclear if the total energy of the jets and isotropic ejecta are the same order of magnitude for all SNe Ic due to the  diversity of SNe's energy reservoirs.
These two should be compatible with either black hole powering or neutron star powering central engines, only if they can power a pair of relativistic jets. Therefore, it is compatible for the magnetar models.
Whether the jets can be produced may depend on the spin and the magnetic field.

In this paper we have proposed a simple full picture of  inner connection for the normal SNe Ic, type Ic super-luminous SNe (SLSNe Ic), X-ray flash (XRF) connected SNe,  broad-line SNe Ic (SN Ic-BL), and gamma-ray burst (GRB) connected SNe. We want to remark that we have selected the interpretations of some observed cases favoring our scenario. In fact there are some other possible interpretations, e.g.  \citet{2011ApJ...728...14P} argued that the type SN Ic-BL, SN 2009bb is not an off-beamed GRB. However, all the different interpretations are far from decisive, and are not always consistent with each other. A selection of some interpretations to provide a logically clear scenario is still helpful to understand the death of the massive stars.


YCZ thanks the helpful discussions with Tsvi Piran, Zigao Dai, Yongfeng Huang, Xiangyu Wang, Zhuo Li, Xuefeng Wu, Yunwei Yu, Fayin Wang, Weihua Lei, Jumpei Takata, Dong Xu, Shanqin Wang, and Wei Xie. Figure \ref{fig-sketch} and  \ref{fig-Miller-dia} are plotted by using Google Presentation and R respectively. We are grateful for the anonymous reviewer for the insightful and detailed comments, and for Prof. Kevin Mackeown for a critical reading of the whole manuscript. KSC is supported by a GRF grant of Hong Kong Government under 17310916. YCZ is supported by the National Basic Research Program of China (973 Program, Grant No. 2014CB845800) and by the National Natural Science Foundation of China (Grant Nos. U1738132 and U1231101).

\nocite*{}

\end{document}